# A Circulating Hydrogen Ultra-High Purification System for the MuCap Experiment

V.A. Ganzha, P.A. Kravtsov[†], O.E. Maev, G.N. Schapkin, G.G. Semenchuk, V. Trofimov, A.A. Vasilyev, M.E. Vznuzdaev,

*Petersburg Nuclear Physics Institute, Gatchina 188350, Russia*

S.M. Clayton, P. Kammel, B. Kiburg,

*University of Illinois at Urbana-Champaign, Urbana, IL 61801, USA*

M. Hildebrandt, C. Petitjean,

*Paul Scherrer Institute, CH-5232 Villigen PSI, Switzerland*

T.I. Banks, B. Lauss

*University of California, Berkeley, and LBNL, Berkeley, CA 94720, USA*

## Abstract

The MuCap experiment is a high-precision measurement of the rate for the basic electroweak process of muon capture, $\mu^- + p \rightarrow n + \nu_\mu$. The experimental approach is based on an active target consisting of a time projection chamber (TPC) operating with pure hydrogen gas. The hydrogen has to be kept extremely pure and at a stable pressure. A Circulating Hydrogen Ultra-high Purification System was designed at the Petersburg Nuclear Physics Institute (PNPI) to continuously clean the hydrogen from impurities. The system is based on an adsorption cryopump to stimulate the hydrogen flow and on a cold adsorbent for the hydrogen cleaning. It was installed at the Paul Scherrer Institute (PSI) in 2004 and performed reliably during three experiment runs. During several months long operating periods the system maintained the hydrogen purity in the detector on the level of 20 ppb for moisture, which is the main contaminant, and of better than 7 ppb and 5 ppb for nitrogen and oxygen, respectively. The pressure inside the TPC was stabilized to within 0.024% of 10 bar at a hydrogen flow rate of 3 standard liters per minute.

[†] Author to whom correspondence should be addressed;

Petersburg Nuclear Physics Institute, 188300, Orlova roscha, Gatchina, Russia;

electronic mail: pkravt@gmail.com. Phones: +7 921 758 9595; +49 0160 160 9292.

## 1. Introduction

The MuCap experiment [1, 2] measures the rate $\Lambda_S$ for the basic electroweak process of nuclear muon capture on the proton, $\mu^- p \rightarrow n + \nu_\mu$, from the singlet hyperfine state of the muonic hydrogen atom. A measurement of $\Lambda_S$ with 1% precision determines the least well-known of the nucleon charged current form factors, the induced pseudoscalar form factor $g_P$, to 7%. The experiment is carried out at the πE3 muon beam line of the 590 MeV proton accelerator at the Paul Scherrer Institute (PSI), Switzerland.

The muon capture rate, $\Lambda_S$, is determined from a comparison of the measured disappearance rate of negative muons in hydrogen, $\lambda^-$, and the world average of the free $\mu^+$ decay rate, $\lambda^+$, by the relation $\lambda^- \approx \lambda^+ + \Lambda_S$. Muons are stopped in a time projection chamber (TPC) with a sensitive volume 15x12x28 $cm^3$ filled with ultrapure hydrogen at 10 bar pressure and room temperature. Muons entering the TPC ionize the hydrogen and the electrons then drift vertically in a homogeneous electrical field to the bottom of the TPC, where they are amplified with a multiwire proportional chamber (MWPC) and read out in two dimensions. In combination with the drift time information, the muon tracks and their stopping points can be reconstructed in three dimensions. This allows a precise identification of muon stops in the TPC and the selection of stop location away from the walls, thus eliminating the background originating from wall stops. The muon decay electrons are then detected in a tracking detector surrounding the pressure vessel, consisting of two wire chambers and a scintillator hodoscope, which covers ~3π solid angle.

Requirements on the hydrogen gas system supporting the experiment are imposed by muon induced atomic and molecular processes taking place in hydrogen and by the operating conditions of the TPC. Negative muons preferentially transfer from μp atoms to heavier elements with high rates. Once a muon has transferred to a $\mu N_Z$ atom, nuclear muon capture proceeds more than 100 times faster than on a μp atom. Thus, even tiny amounts of impurities in the hydrogen gas distort the observed μp lifetime spectrum of interest. Consequently, muon transfer and subsequent capture must be suppressed by keeping the gas contaminants below a level of typically 10 ppb.

Additionally, isotopically pure hydrogen is required, since muons transfer to deuterium, where they pose a systematic problem for the MuCap experiment. Due to a Ramsauer-Townsend minimum in the μd + p scattering cross-section [3], μd atoms can diffuse over macroscopic distances and can either escape from the stopping volume in the TPC in a time-dependent way and can even reach the chamber materials, where muons are quickly captured. A dedicated isotope separation device [4] was built to produce deuterium depleted hydrogen ("protium") with a deuterium contamination of less than 70 ppb. This work will be described in a forthcoming paper.

As the gas amplification in the MWPC of the TPC sensitively depends on the hydrogen density, the pressure inside the TPC must be stabilized on the level of 10 bar with 0.1% accuracy to guarantee stable running conditions, namely to keep the MWPC gas gain constant within 1%. There is also concern that large flow variations in the TPC might induce dust accumulation on the TPC electrodes leading to breakdown of the chamber high voltage. Thus the flow rate must be stabilized as well. Finally, reliability and hydrogen safety were of utmost importance as the whole detection system operated with high voltage up to 30 kV in a pure hydrogen environment.

Table 1 summarizes the main design criteria of the gas system. The "Circulating Hydrogen Ultrahigh Purification System" (CHUPS) was designed and built to provide continuous protium purification. CHUPS achieved or surpassed most design criteria over several experimental runs with typically two months of continuous operation per year.

| **Parameter** | **Required** | **Achieved** |
|---|---|---|
| Operating pressure | 10.0 bar | 10.0 bar |
| Pressure variations | <0.1 % | 0.024 % |
| Flux | 0.5-1 slpm | 0.5-3 slpm |
| Flux stability | 0.5 slpm | 0.08 slpm |
| $N_2$ contamination | ≤10 ppb | 7±1 ppb |
| $O_2$ contamination | ≤10 ppb | <5 ppb |
| $H_2O$ contamination in the TPC | ≤10 ppb | 18 ppb |
| $H_2O$ content at the CHUPS outlet | <10 ppb | ~5 ppb |
| Cleaning capacity[1] | 20 mg | 1000 mg |
| Typical operation period | 2-3 months | 2-3 months |

Table 1. Comparison of the main parameters relevant to the design of CHUPS and the values achieved during operation. Flux units are given in standard liters per minute (slpm).

## 2. Impurity accumulation in the TPC

The TPC is mounted in a 40-liter cylindrical pressure vessel, filled with 10 bar hydrogen gas at room temperature. Only high vacuum-proof and bakeable materials were used inside the pressure vessel. The hydrogen vessel and TPC detector were baked under vacuum at 115 °C for extended time periods (weeks to months). The hydrogen was then filled through a palladium filter[2] which can provide an initial purity of better than 1 ppb. However, it is impossible to maintain this high purity level in a large vessel containing a complicated device over several weeks. Admixtures are continuously accumulated from various sources, specifically from diffusion through bearings, and outgassing from the inner surface of the hydrogen vessel, the TPC materials and the surfaces of connected gas lines. The most abundant impurity is generally moisture, formed as hundreds to thousands of monomolecular layers [5] on any surface while exposed to air. These layers can be removed only by continuous pumping or flushing with very dry gas, combined with baking at 450 °C. As the TPC materials allowed heating only up to a maximum temperature of 120 °C, the humidity was significantly reduced, but not eliminated.

[1] With respect to water as the main impurity.

As will be described in Section 4.2, the TPC can directly monitor the muon capture reactions on impurities, with a resulting impurity capture event yield proportional to the impurity concentration. A plot of impurity capture events (Fig. 1), measured before CHUPS installation, shows the impurity capture event yield increasing in the hours after filling the hydrogen vessel through the palladium filter. The figure corresponds to an impurity accumulation rate of tens of ppb per day, depending on the molecular species, which is much larger than required for the experiment. This was the main reason for the development of CHUPS, designed for continuous gas cleaning and circulation through the operating TPC. The impurity accumulation was also confirmed with chromatographic measurements.

---

[2] 8 slpm Hydrogen Purifier. REB Research & Consulting. Ferndale, MI 48220, USA (http://www.rebresearch.com).

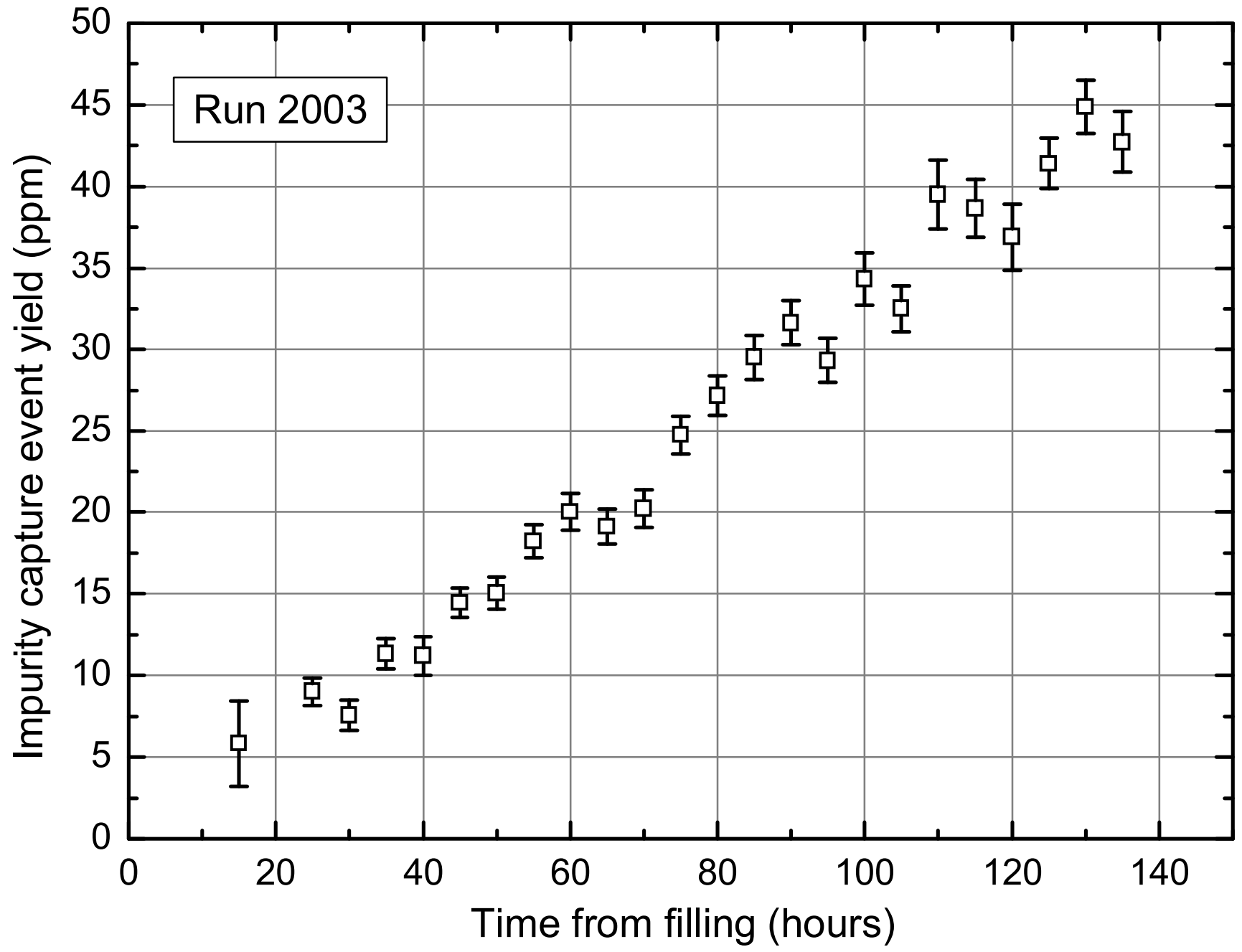

Fig. 1. Accumulation of contaminants in the hydrogen vessel monitored by the observation of impurity capture events in the TPC.

## 3. Design of the circulating system (CHUPS)

### 3.1. CHUPS operating scheme

Consider a circulating system of volume V connected to the TPC detector. The circulation flow of the system (Q) has the residual concentration $C_0$ of impurity contaminations. The hydrogen gas in the TPC vessel contains impurities with concentration $C_z$ and the vessel's outgassing flow is $Q_z$, measured in [ppm·slpm]. The impurity mass balance can then be written as differential equation:

$$V\frac{dC_z}{dt} = -Q(C_z - C_0) + Q_z .$$

The solution of this equation is:

$$C_z = C_0 + \frac{Q_z}{Q} + Ae^{-\frac{Q}{V}t},$$

where coefficient A is defined by the initial conditions. The equilibrium partial pressure after long term purification is

$$C_z = C_0 + \frac{Q_z}{Q}.$$

The residual impurity concentration at the circulation system outlet was measured to be small and can be neglected. Thus, the impurity concentration in the TPC detector is defined by the ratio of outgassing flow and circulation flow.

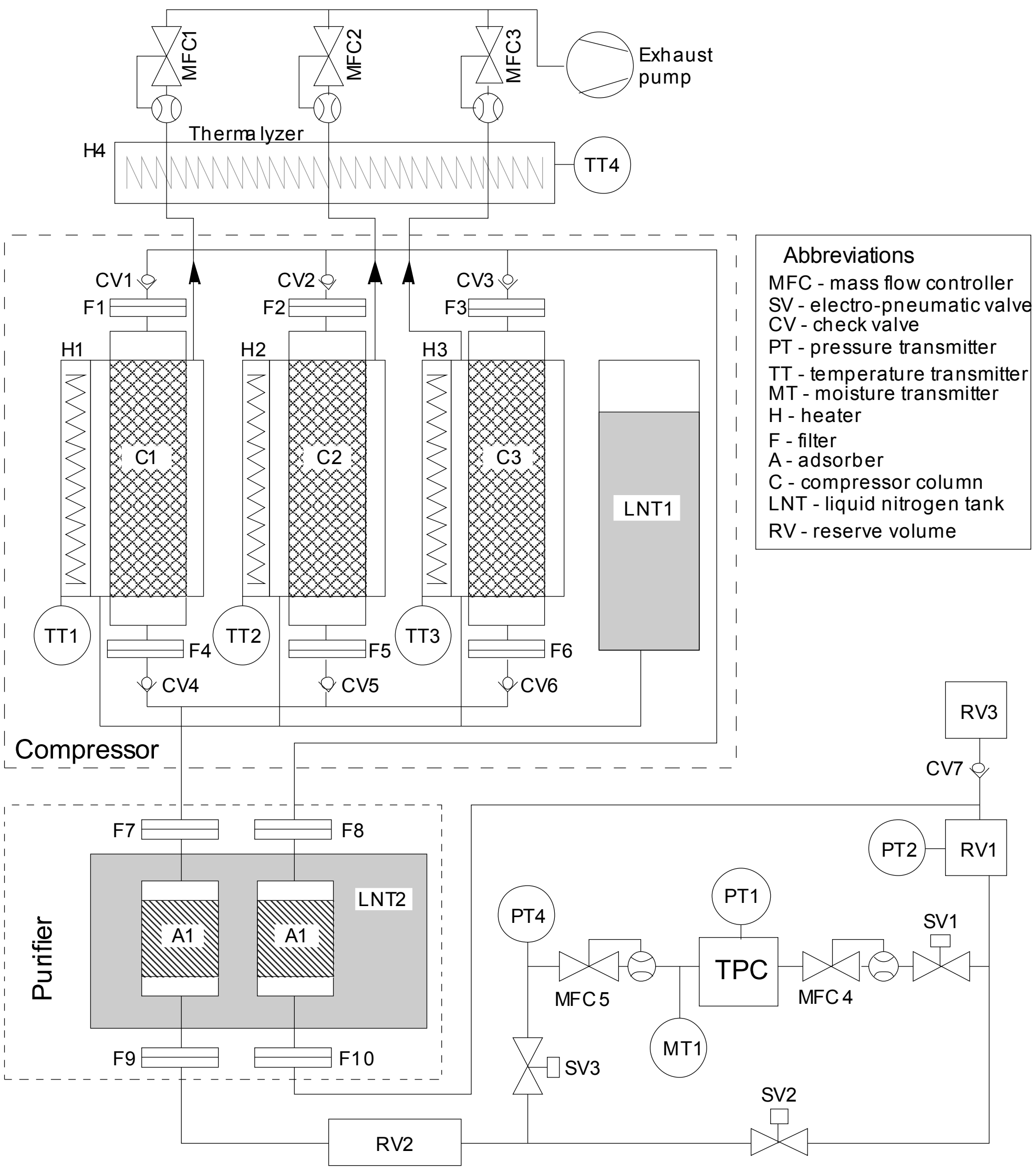

Fig. 2. Simplified CHUPS diagram.

Initially a circulation system consisting of a mechanical pump and a palladium (Pd) filter was considered [1]. High vacuum rated pumps that operate at 10 bar pressure are not commercially available, and Pd filters supporting the required flux of 3 slpm are expensive. Instead, the alternative CHUPS scheme was realized, which is based on an adsorption cryopump to maintain the hydrogen flow and a cryogenic adsorption filter for removing the impurities. The adsorption cryopump has essential advantages, such as intrinsic high purity and reliability due to the absence of moving parts. The circulating system was designed, prototyped,built and tested at the Petersburg Nuclear Physics Institute (PNPI) in Gatchina, Russia, and installed at PSI during the preparation period of the MuCap experiment in 2004 [6]. It was upgraded with a humidity monitoring system and additional flow stabillization in 2005. A simplified diagram of the system is shown in Fig. 2. CHUPS consists of two main units which are mounted separately on a common frame: compressor and purifier.

### 3.2. Compressor

The compressor is a triplex adsorption cryopump. It has three identical cartridges (columns) filled with activated carbon[3] and connected in parallel. The cartridges are made from thick-walled stainless steel tube of 38 mm diameter and designed to keep high pressure. The volume of each column is about 1 liter and contains up to 0.6 kg of the adsorbent. Each column (C1, C2, C3 in the scheme) has a heat exchanger made of copper tube. It is coiled around the column and soldered to its outer surface by silver hard alloy. The heat exchanger is used to cool down the compressor column using liquid nitrogen flowing from the 40-liter supply vessel (LNT1). Liquid nitrogen flow is provided by an exhaust pump connected to the heat exchangers' manifold. After cooling the columns, nitrogen flows through the thermalizer (H4), which heats the gas to room temperature. Mass-flow controllers[4] MFC1, MFC2, MFC3 control the flow rate of the gaseous nitrogen and consequently regulate the liquid nitrogen flow rate and the cooling rate of the column. An electric heater is coiled around the column between the heat exchanger turns to provide the column heating. Each column has two check valves with an actuation pressure of 50 mbar that are installed in the inlet and outlet pipelines (CV1-CV6). Inlet and outlet lines of the columns are combined to the inlet and the outlet manifolds, respectively.

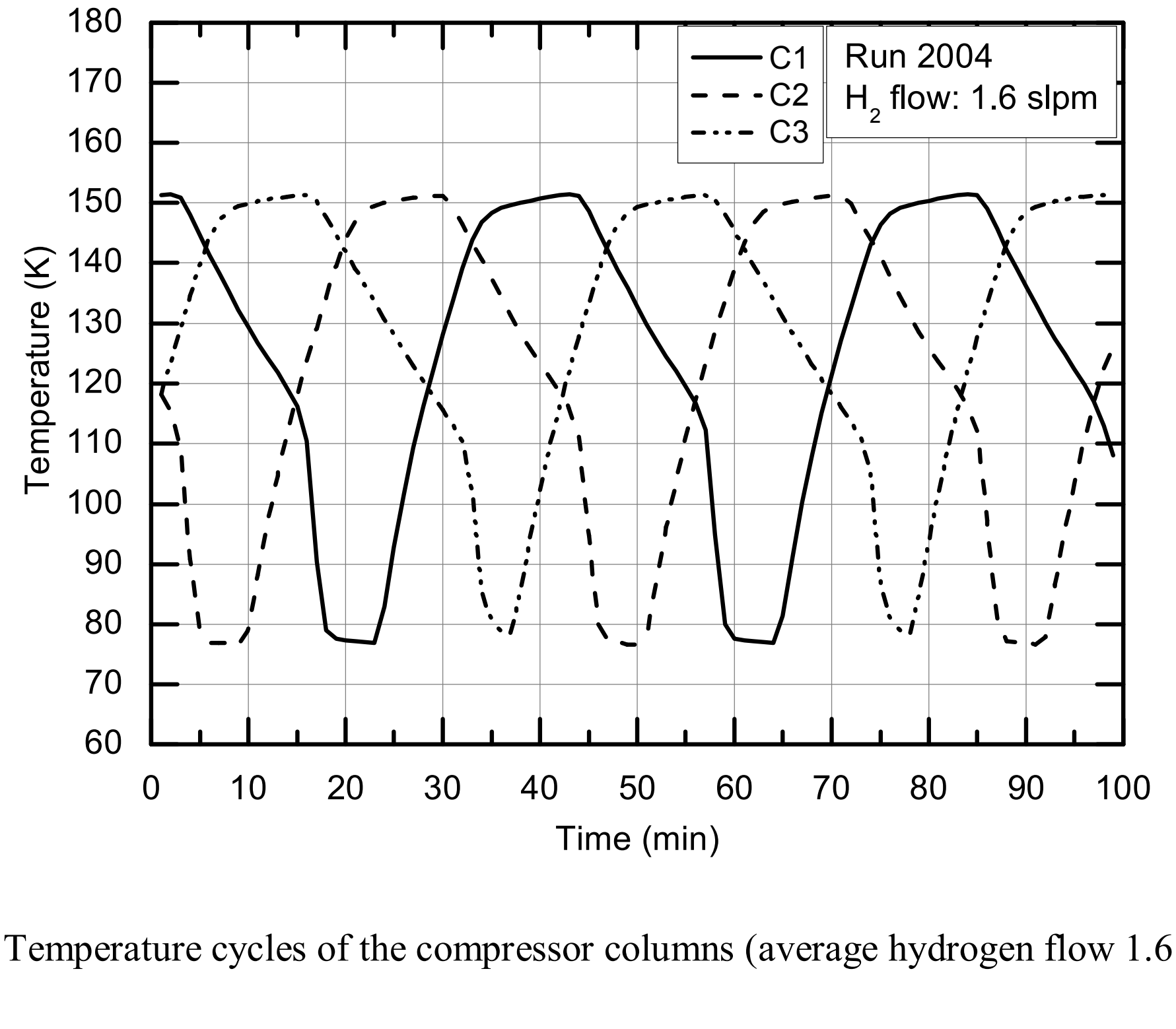

Fig. 3. Temperature cycles of the compressor columns (average hydrogen flow 1.6 slpm).

---

[3] Norit Nederland B.V. Nijverheidsweg Noord 72, 3800 AC AMERSFOORT, The Netherlands.

[4] Stainless steel GFC series mass-flow controllers. Aalborg, 20 Corporate Drive, Orangeburg, New York, 10962, USA. (http://www.aalborg.com/)

At the cooling stage the adsorbent inside the column adsorbs hydrogen, and the internal pressure drops below the pressure in the inlet line of the compressor. Consequently, the inlet check valve opens and passes hydrogen into the column. During the heating stage the adsorbent desorbs hydrogen, and the column internal pressure rises above the outlet line pressure. The outlet check valve opens and passes hydrogen into the outlet line. Thus, the combination of the check valves provides a pulsating flux of hydrogen in one direction. Cooling and heating rates are regulated by the balance of liquid nitrogen flow (controlled by MFC1, MFC2, MFC3) and heating power (managed by pulse-width modulation of the power supplies).
The temperature phases of the columns are shifted with respect to each other as shown in Fig. 3. Upper and lower temperatures and cycle frequency are regulated in accordance with the required average flow rate. A single compressor column pumps approximately 32 standard liters of hydrogen in one cycle employing a temperature range of 80-150 K. The maximum hydrogen flow of 3 slpm through the compressor is defined by the maximum cooling rate and the thermalization time of the activated carbon.

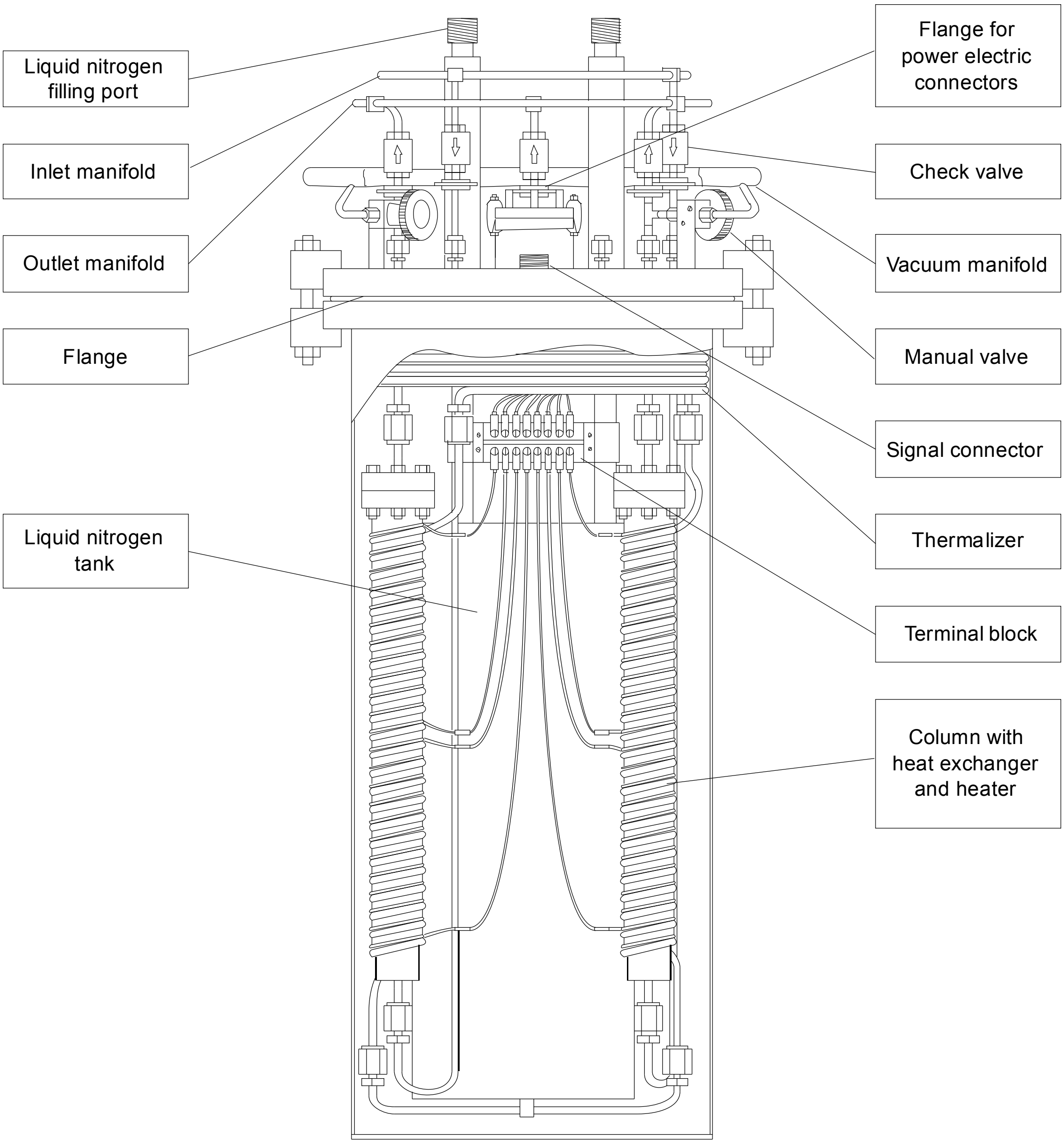

Fig. 4. Compressor layout.

Fig. 4 shows the detailed compressor layout. The columns are arranged around the liquid nitrogen tank, all contained in a vacuum case. The capacity of the 40-liter liquid nitrogen tank allows CHUPS to operate for eight hours at the maximal available flux without refilling of liquid nitrogen. The tank is filled through vacuum insulated filling ports. The check-valves are distributed outside of the vacuum case and connected with inlet and outlet manifolds. The vacuum manifold provides evacuation of the columns during their regeneration. It is connected to the inner space of each column by manual valves.

All electric lines of the compressor are connected to a terminal block and come out of the vacuum case through a separate small flange with vacuum tight signal connectors. The thermalizer is fixed under the vacuum case flange. It is made of the same copper tubes as the heat exchangers in order to prevent ingress of cold nitrogen into mass flow controllers.

### 3.3. Purifier

The purifier contains two cartridges filled with NaX-type zeolite[5] (indicated as adsorbers A1 and A2 in Fig. 2). The adsorbers are immersed into liquid nitrogen in a separate vessel (LNT2) and permanently kept under a temperature of 77 K during the entire MuCap experimental run. Low temperature is essential to increase the adsorption ability of zeolite for high-boiling contaminants (oxygen, nitrogen) against the main gas (hydrogen). The two adsorbers contain about 40 g of the sorbent in total, enough to accumulate up to 1 g of adsorbed water. The total amount of water supplied by the flow of 3 slpm during two months was about $2 \cdot 10^{-2}$ g (allowing for 100 ppb constant humidity), which is 2% of the adsorber capacity. The adsorption capacity for other contaminants is comparable. Thus, the adsorbers guarantee full impurity removal in the hydrogen flow during the long term experiment. Before each experiment run, the adsorbent has to be exchanged or can be regenerated by heating up to 400 °C and pumping.

The liquid nitrogen vessel is contained in the vacuum case and protected from external thermal radiation by a copper shield mounted on the secondary liquid nitrogen vessel, which is also used to cool down the incoming hydrogen. This technique decreases the liquid nitrogen consumption and prevents heating of the zeolite adsorbers. The two adsorbers of the purifier (A1 and A2) are mounted on the hydrogen lines upstream and downstream of the TPC, respectively, providing two-stage purification of the gas. The inlet zeolite adsorber removes most of the impurities, which helps to avoid any decrease of the compressor capacity due to accumulation of impurities in the activated carbon.

The system is equipped with mechanical 2μm mesh filters installed before and after the compressor and in the detector pipelines that prevent carbon or zeolite dust penetration to the clean part of the system. The final purification is provided by a special gas purifier[6] of limited capacity installed at the TPC inlet.

### 3.4. Control system

A special stand-alone microcontroller block provides all necessary regulation algorithms. The control block is connected to a PC via RS-232 or RS-485 serial interfaces. By implementing all control algorithms in an independent control block, the system remains operational in case of computer failure.

Computer software is used for adjusting the parameters of the regulation algorithms, collecting and visualizing the process variables and keeping them in the database. All system events (including software messages and alarms) are also saved in the database. Custom software was developed to access the parameters history and event log in the database.

---

[5] CECA company, 89 Boulevard National, La Garenne Colombes, 92257, France. (http://www.adsorbents.com/sites/ceca/en/home.page)

[6] SAES Pure Gas - MicroTorr Ambient Temperature Gas Purifiers. SAES Pure Gas, Inc. 4175 Santa Fe Road, San Luis Obispo, California, 93401, USA. (http://www.puregastechnologies.com/microt.htm)

The control block measures and controls all system devices such as pressure and temperature sensors, valves, mass-flow controllers, etc. It provides the following regulation procedures:

- temperature stabilization of the three compressor columns and the thermalizer by regulating the heating and cooling provided by the heaters and the nitrogen mass-flow controllers, respectively;
- cyclic operation of the compressor columns with phase shift;
- TPC internal pressure stabilization using mass-flow controllers;
- temperature stabilization of the humidity sensor.

Alarm and interlock functions are also implemented in the control block firmware (Table 2). It protects the TPC detector from underpressure and overpressure and controls the differential pressure between the TPC and reserve volume RV1 to avoid hydrogen flux variations. The software also tracks the liquid nitrogen level in the compressor tank (LNT1) and the compressed air pressure that is used for the electro-pneumatic valves[7] (SV1-SV3) actuation. All alarm events triggered a light and sound signal.

| **Alarm description** | **Condition** | **Action** |
|---|---|---|
| TPC pressure high | PT1>PT1max | Cut-off TPC and open bypass |
| TPC pressure low | PT1<PT1min | Cut-off TPC and open bypass |
| Differential pressure high | (PT2-PT1)>DPmax | Cut-off TPC and open bypass |
| Differential pressure low | (PT2-PT1)<DPmin | Cut-off TPC and open bypass |
| Compressed air pressure low | PT3<PT3min | Alarm signaling |
| Liquid nitrogen level low in LNT1 | N2Level<N2Levelmin | Alarm signaling |

Table 2. Alarm events and interlock actions.

The passive safety system consists of two relieve valves. Valve (CV7) set to 13 bar is installed in the main CHUPS circulation circuit to release protium gas into a 30 liter reserve volume (RV3) in case of dangerous overpressures (see Fig. 2). The TPC vessel has a separate relieve valve set at 12 bar to protect the 0.5 mm thick beryllium beam entrance window.

### 3.5. TPC pressure and flow control

The three columns of the compressor induce the hydrogen flow to exchange the gas in the TPC. This flow is pulsating because of the periodical mode of the column operation. The pressure inside the TPC must be stabilized with a 0.1% accuracy. In order to smooth the pressure variations caused by the compressor, the CHUPS system is equipped with two reserve volumes of 15 liters each (RV1 and RV2), pressure sensors[8] (PT1, PT2, PT4) and mass-flow controllers (MFC4 and MFC5).

[7] Type SS-6BK-MM-1C valves, Swagelok company, 29500 Solon Road, Solon OH 44139, USA.

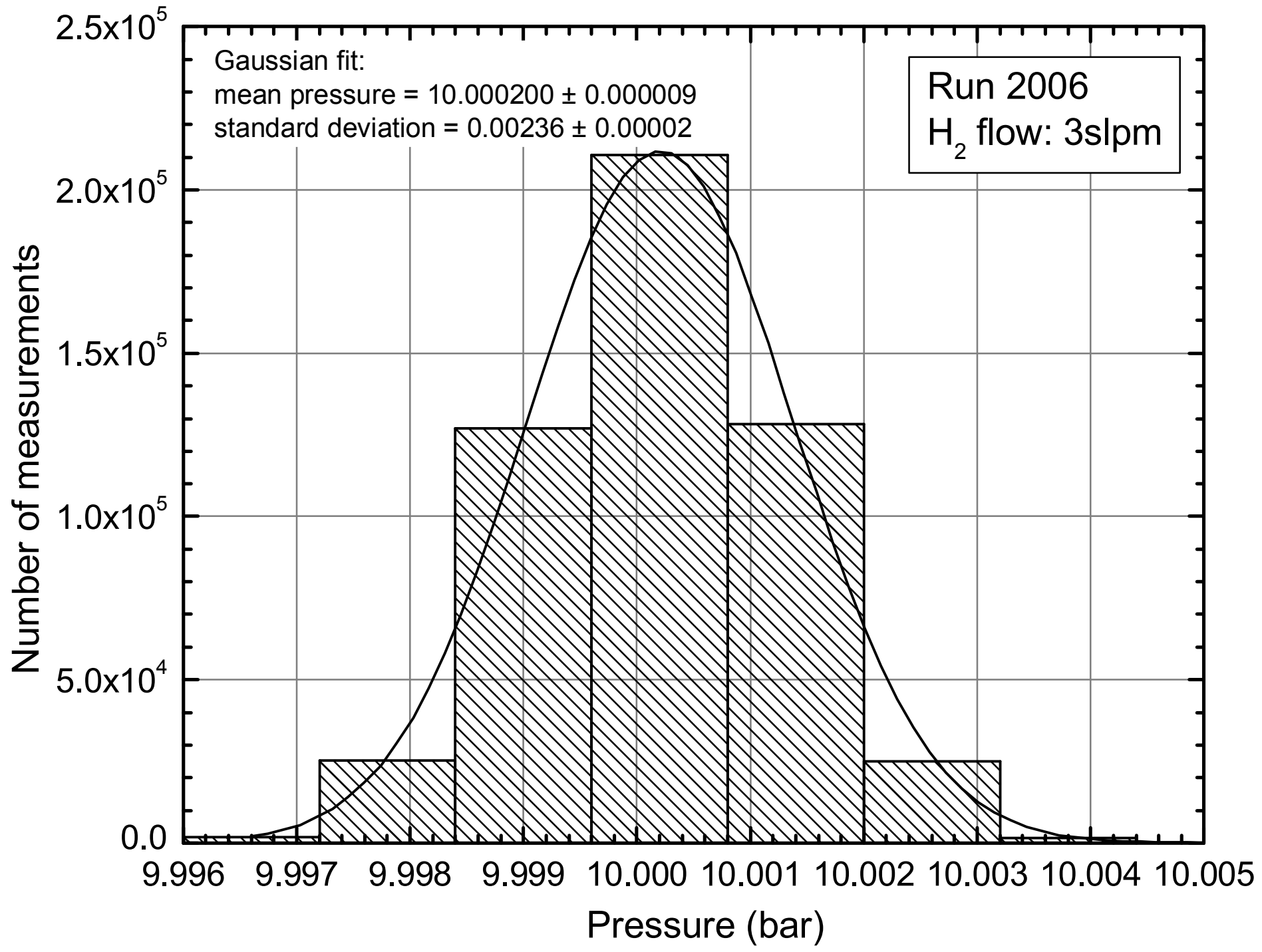

Fig. 5. TPC pressure distribution over 50 days. The gaussian fit has a standard deviation of $\sigma = 2.4$ mbar representing a relative stability of 0.024%. The absolute precision is limited by the absolute calibration error of the sensor (0.1%).

The reserve volumes are installed in the inlet and outlet lines of the TPC. They are used as buffering volumes. RV1 also provides a hydrogen reserve to support the pressure stabilization algorithm. Each volume is equipped with a pressure sensor (PT2 and PT4, same model as PT1). Two mass-flow controllers[9] with a maximal flow of 20 slpm are mounted at the inlet (MFC4) and outlet (MFC5) lines of the TPC. The internal pressure of the detector (measured by PT1) is stabilized using Proportional-Integral-Derivative (PID) regulation. The mass-flow controller at the detector outlet (MFC5) is set to a constant flow rate. The TPC inlet mass-flow controller (MFC4) is operated by PID algorithm in the control software. The MFC5 set point defines the average flow rate through the TPC vessel. The pressure distribution histogram (Fig. 5) for the long term operation during 50 days shows excellent relative stability of 0.024% at a mean hydrogen flow of 3 slpm. The histogram binning corresponds to the ADC digitization in the control system (1.2 mbar). Thus the pressure is kept within ±2 least significant bits of the 16-bit

---

[8] Keller Piezoresistive pressure transmitters series PAA21. KELLER AG für Druckmesstechnik, St. Gallerstrasse 119 CH-8404 Winterthur.

[9] Brooks smart mass flow controllers model 5850S. Brooks Instrument, 407 West Vine Street, Hatfield, PA, 19440-0903, USA. (http://www.emersonprocess.com/brooks/).

ADC. The second reserve volume (RV2) and the mass flow controller (MFC5) at the compressor inlet were installed to prevent pressure drops caused by the fast opening of the check valve. This resulted in a flow stability of 2.5% at the mean flow level of 3 slpm during 50 days of CHUPS operation (Fig. 6).

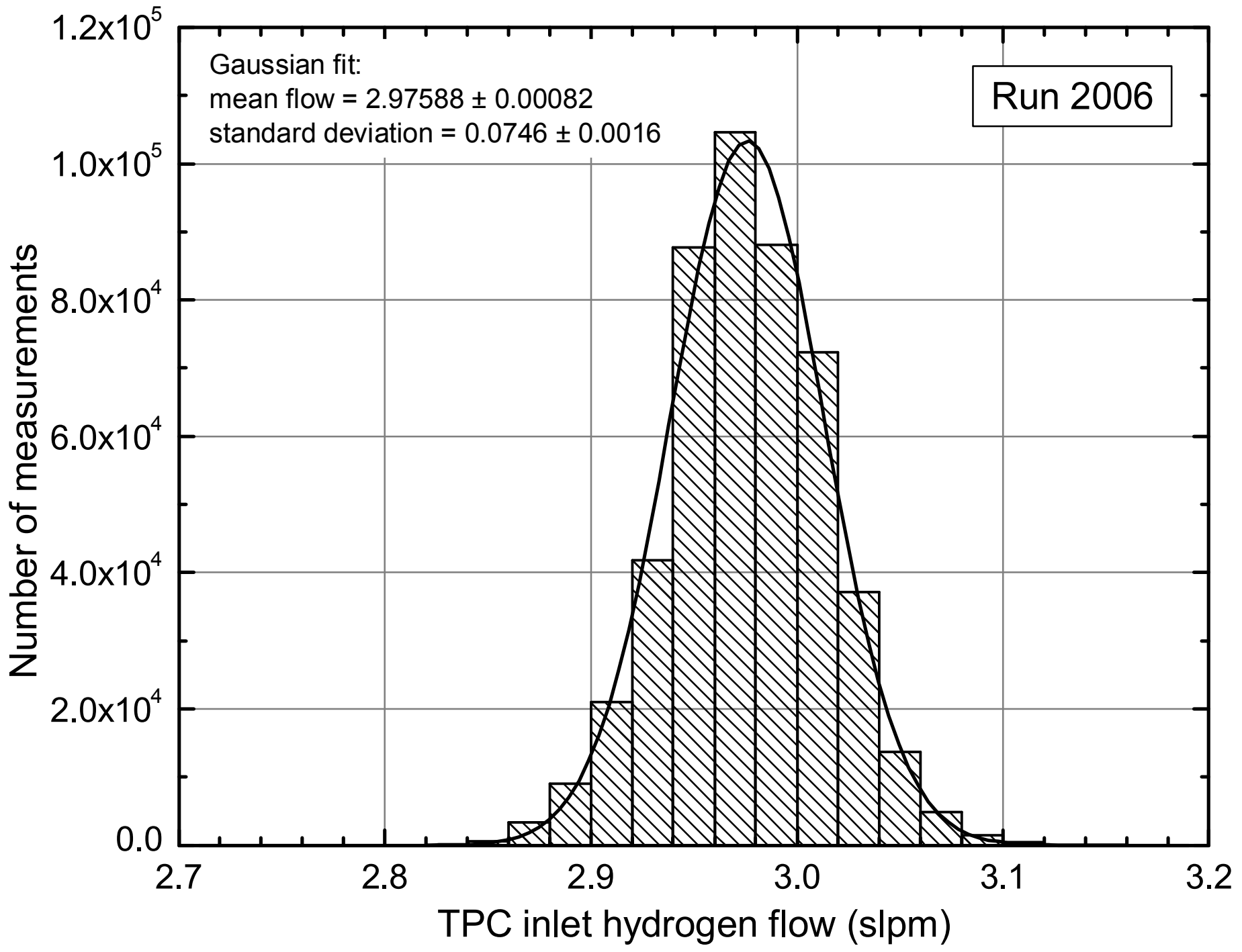

Fig. 6. TPC inlet flow distribution over 50 days. The gaussian fit has a standard deviation of $\sigma = 0.075$ slpm representing a relative stability of 2.5%. The absolute precision is limited by the absolute calibration error of the mass flow controller (2%).

## 4. Monitoring impurities

The initial purity of the evacuated TPC system was achieved by continuous pumping and baking of the TPC vessel prior to the experiment. The residual gas contents were controlled by a quadrupole mass spectrometer in the mass range from 1 to 100 atomic mass units (a.m.u.). Throughout the accumulation of the production MuCap data, several samples of the hydrogen gas were taken and the quantities of atmospheric gases (oxygen and nitrogen) were determined offline by sensitive gas chromatography. The humidity in the gas was monitored online with a dew-point transmitter[10]. The total amount of impurities was continuously monitored by the observed impurity capture event yield in the TPC.

---

[10] PURA High Purity Gas Dew-point Transmitter. The Kahn companies, 885 Wells Road Wethersfield, CT 06109 USA. (http://www.kahn.com/)

### 4.1. Chromatographic analysis of nitrogen and oxygen

The oxygen and nitrogen content is measured by gas chromatography with a thermal conductivity sensor in the setup shown in Fig. 7. After being filtered and purified in a cryogenic adsorption purifier, helium carrier gas is distributed in two directions with equally adjusted flow rates. The first flow passes through the reference chromatographic column. The second flow is either routed to the working chromatographic column directly or through the accumulating column using the volume batcher that injects a fixed volume into the specified direction. Both chromatographic columns are filled with a specially treated adsorbent (zeolite). The adsorbent in the working column separates the admixtures, while the one in the reference column provides a hydraulic resistance equal to the working column.

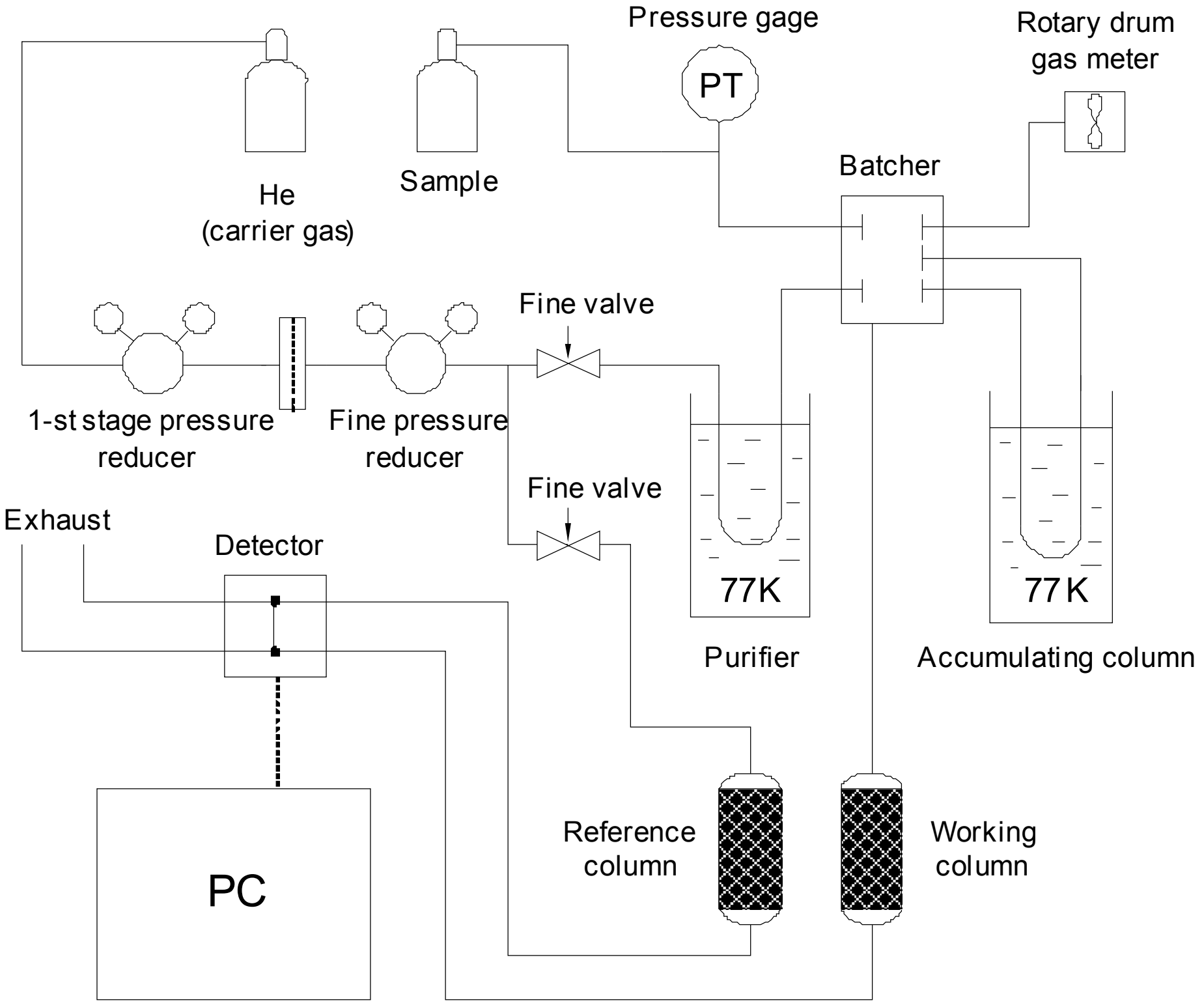

Fig. 7. Layout of the gas chromatographic setup.

Preparatory enrichment is required for the very low concentrations of impurities in the hydrogen gas samples. Hydrogen from the sample bottle is directed to the accumulating column by the batcher. This column is filled by zeolite and immersed into a liquid nitrogen vessel at 77 K temperature. A rotary drum gas meter measures the quantity of gas passed through the accumulating column. Impurities from the gas stream are adsorbed by the zeolite and remain in the accumulating column. After the desired amount of sample gas has passed through the adsorber, the accumulating column is removed from the cryogenic vessel and heated. Then the batcher directs the He carrier gas flow through the accumulating column, which washes the impurities out of the adsorbent.

In the next step, the carrier gas enriched by contaminants flows through the working column, where contaminants are separated on the adsorbent. Both reference and working flows pass in parallel through the detector, which measures their differential heat conductivity. The differential output signal is proportional to the admixture concentration. Signal peaks are registered by a PC and processed by cuts in software[11]. The enrichment coefficient (the sample volume passed through the accumulating column) and parameters of the measuring scheme are adjusted with respect to the admixture concentration. The final calculation of the concentration is based on a calibration measurement. A serial dilution method was used to obtain the set of calibrating samples with decreasing concentration of nitrogen and oxygen. With this method a sample of air was serially attenuated by mixing with hydrogen. The total calibration error, including non-linearity, is 10%.

The traces of oxygen and nitrogen were monitored by gas samples of 8-10 normal liters from the TPC operating gas using the chromatographic method. A sensitivity of 5 ppb for oxygen and 7 ppb for nitrogen was established. Fig. 8 shows an example of chromatographic measurements of commercial hydrogen (10 ppm of oxygen and 31 ppm of nitrogen) and for a sample from the TPC output during the regular clean operation. The measurement of this clean sample indicates 10±1 ppb of nitrogen and no oxygen traces at the sensitivity level of 5 ppb, which is confirmed by the same peak positions for the clean and the commercial samples (indicated by dashed vertical lines).

[11] Chromprocessor. Software and hardware for gas chromatography. http://user.rol.ru/~zapisnyh/cprekl.htm.

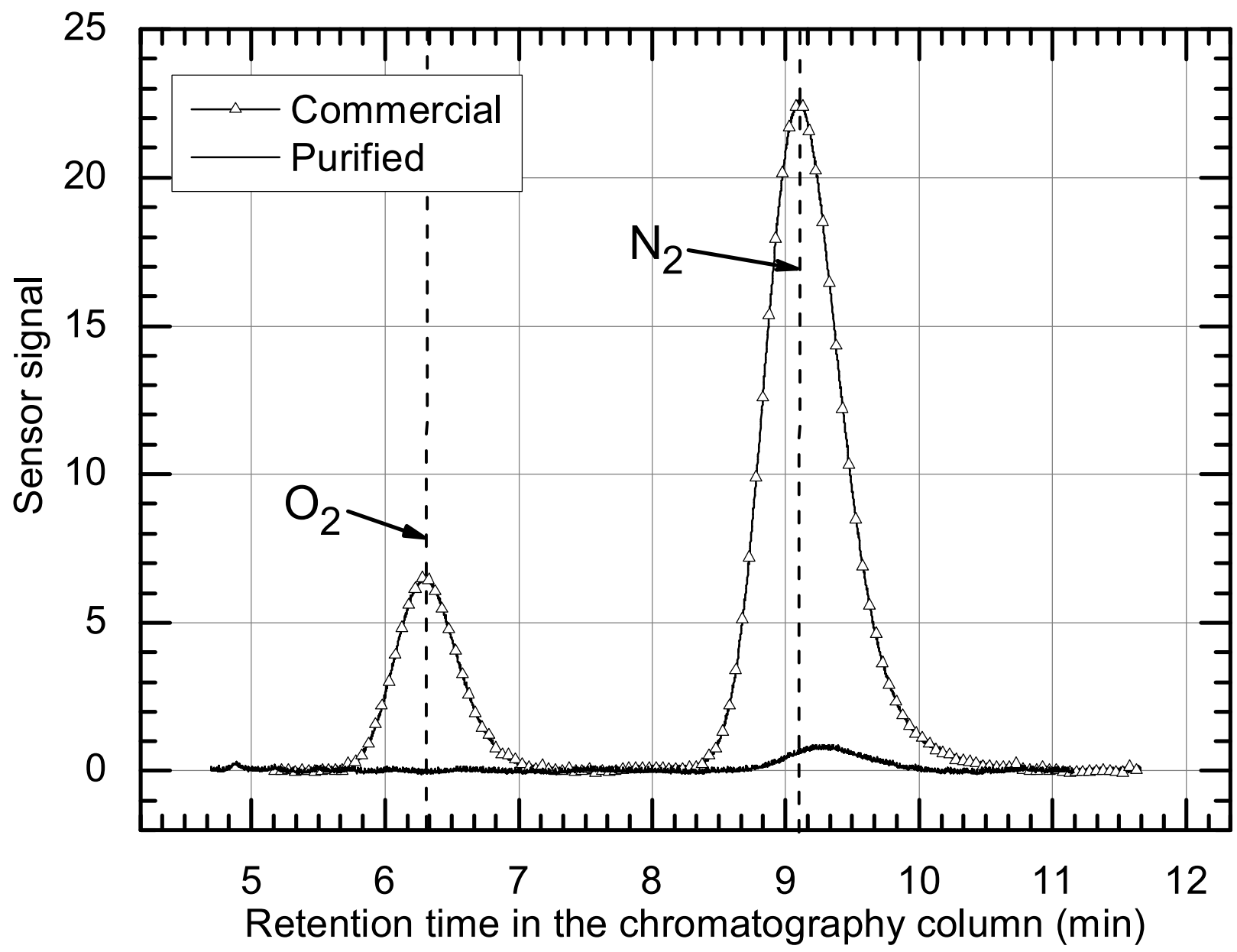

Fig. 8. Commercial and purified hydrogen chromatograms. Positions of expected oxygen and nitrogen peaks are indicated.

### 4.2. Muon capture on impurities monitored via capture event detection in the TPC

One of the unique capabilities of the MuCap experiment is an online in-situ measurement of ppb impurity concentrations in the hydrogen gas. At MuCap hydrogen gas conditions the vast majority of negative muons remain in muonic hydrogen atoms ($\mu p$) and only approximately 5% form muonic molecular states ($p\mu p$) before decay. However, a small fraction of muons, proportional to the concentration $c_Z$ of a contaminating $Z>1$ impurity, can transfer to this impurity and then capture on the nucleus with a much larger rate $\Lambda_Z$ as compared to the capture process on the proton. Omitting small bound state corrections, the nuclear muon capture reaction occurs with a probability proportional to $\Lambda_Z/(\Lambda_Z+\lambda^+)$, as a result of the competition between muon capture and muon decay. The capture process $\mu N_Z \rightarrow N_{Z-1} + \nu$ is identified by its distinct signature in the TPC as shown in Fig. 9. The energy deposition of a beam muon in the TPC follows a Bragg curve, i.e. the muon deposits the largest energies close to the anode where it stops. In case of an impurity capture such a muon track is followed in time by a large signal from the $N_{Z-1}$ recoil nucleus confined to the close vicinity of the stop anode. This sequence is unique and allows monitoring of impurities at the ppb level.

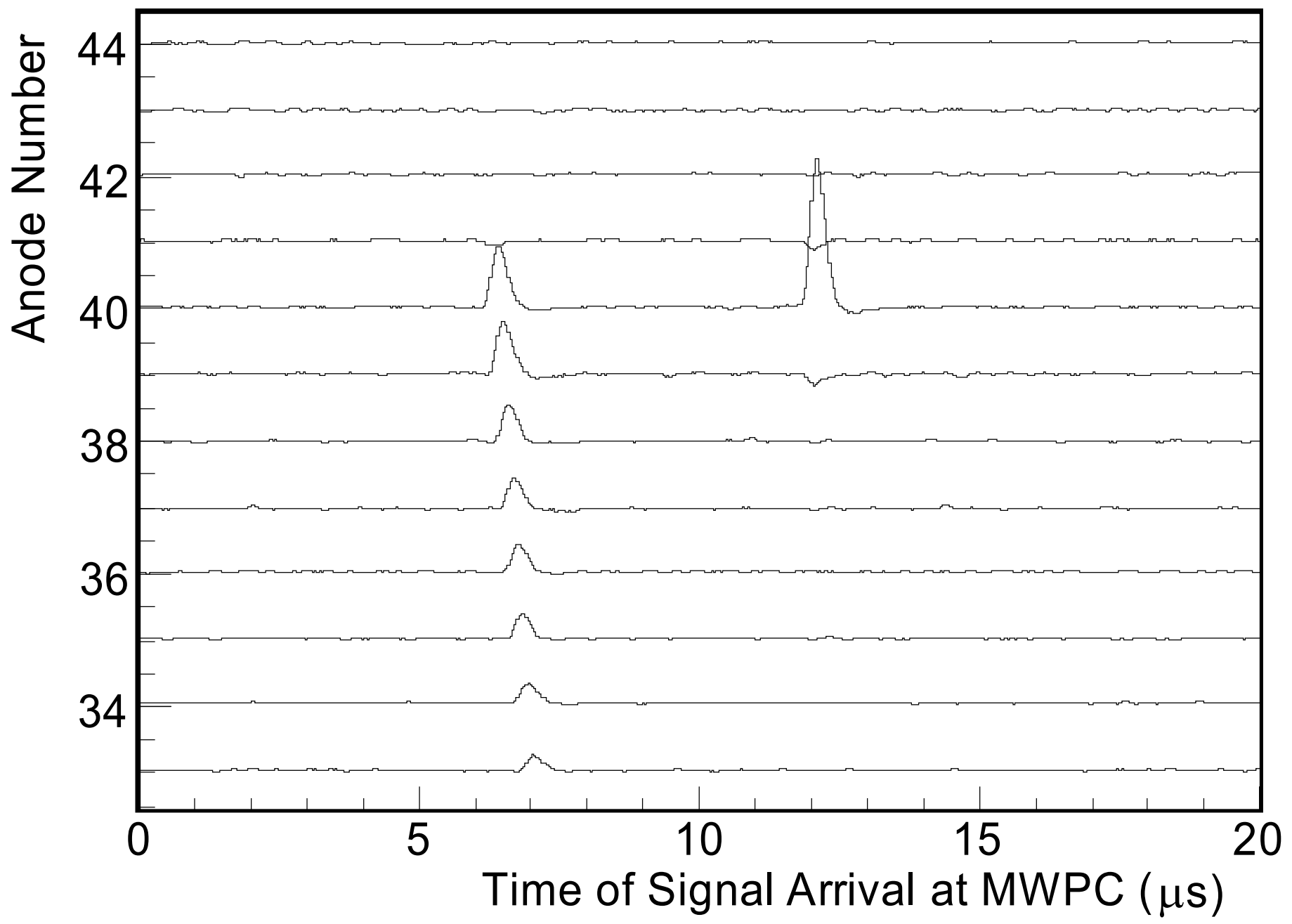

Fig. 9. Signature of a capture event in the TPC. The incident muon stops on anode 40 and is followed by a large, local signal on the stopping anode generated by the recoiling $N_{Z-1}$ nucleus.

The observed yield of such capture events per stopped muon can be written as $Y_Z = c_Z\ k_Z$, where $k_Z$ are coefficients which depend on the transfer rate, capture rate $\Lambda_Z$ and detection efficiency for each contaminant. The coefficients were determined in dedicated calibration runs where the hydrogen gas was doped with a single known impurity component of well measured molecular concentration $c_Z$, 100-1000 times higher than in the clean run conditions. Typical measured values were $k_N$=70 (for nitrogen) and $k_{H2O}$=200 (for $H_2O$) where both $Y_Z$ and $c_Z$ are given in ppm. The detection efficiencies vary only slightly from run to run, depending on the exact signal threshold settings.

In summary, the capture yield measurement allows a continuous, precise and very sensitive measurement of the overall observed capture yield from impurities. The possibility of determining the elemental impurity compositions from the energy spectra of the capture recoils is currently being studied.

**4.3. Analysis of humidity concentration in hydrogen**

In 2004, we observed a capture yield of ~10 ppm which would correspond to about 140 ppb of nitrogen, significantly higher than the oxygen and nitrogen concentrations derived from the gas chromatography. This pointed towards an additional contaminant that was not observed in the gas-chromatographic method.

Therefore a dew-point transmitter for online humidity measurements was installed in the next run (2005) to determine whether the additional observed yield can be explained by residual water vapor in the hydrogen gas, which was the suspected contaminant. The sensor has a sensitivity of 0.02 ppb with a calibration error of +30 / -50 % in the 2-100 ppb range. It was mounted in a temperature-stabilized box in order to reduce any influence of the ambient temperature to the sensor reading. The sensor and inlet pipeline were kept at 21 °C with 0.2 °C stability. The humidity sensor (MT1) was installed in the gas circuit such that it could measure either the humidity in the outlet TPC flow or in the isolated CHUPS system (while hydrogen is circulating through a chamber bypass). Continuous circulation with the TPC volume bypassed gave 5 ppb humidity. During the production run the humidity sensor clearly demonstrated that humidity was the main additional impurity, explaining the discrepancy between gas-chromatography measurements and the impurity capture event yield observed in the TPC.

## 5. CHUPS operating experience

### 5.1. Clean hydrogen fills for main physics data

Before the 2004 experiment the TPC was baked for several weeks. The performance of the gas system during the run period is documented in Fig. 10. At t=0, CHUPS operation starts and the impurity capture event yield drops significantly until an equilibrium state is reached and the yield levels off. During a short disconnection of CHUPS from the hydrogen vessel from hour 60-80, outgassing inside the vessel becomes immediately visible. The reconnection of CHUPS around hour 80 again cleans the gas. During the time of reduced 1 slpm operation (Fig. 10c), the outgassing was faster than the cleaning power of the CHUPS system and the corresponding impurity increase can be observed between hours 340 to 360 in Fig. 10b. Finally, the yield levels off at the equilibrium value between CHUPS cleaning and outgassing with a flow rate of 2.6 slpm.

The saturation level of the capture yield data is slightly decreasing for the 1.6 and 2.6 slpm flow. For this reason the hydrogen flow was increased to the maximal level of 3 slpm in the follow-up measurements (2005-2006).

Chromatography measurements also track the purification process (Fig. 10a). Due to limited capacity of the circulation system and the detector, only few probes were taken. Oxygen traces dropped below the apparatus' sensitivity of 5 ppb within two days after starting the circulation. The final chromatographic measurements resulted in nitrogen concentrations below the apparatus' sensitivity of 7 ppb.

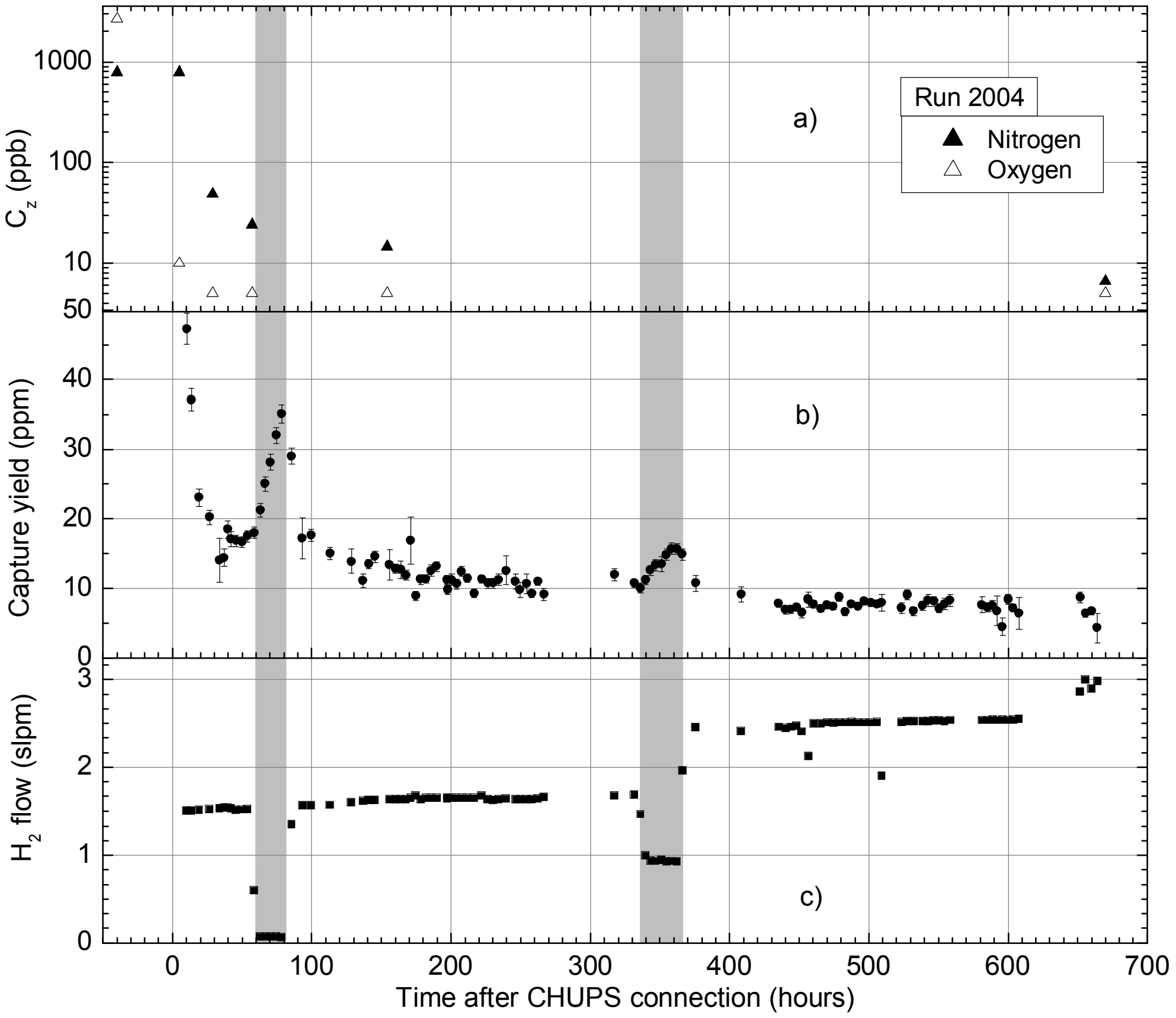

Fig. 10. Monitored impurity concentrations during the MuCap production run in 2004.

a) Results of gas-chromatographic analysis of operating gas samples. The values for nitrogen track the observed impurities behavior, but cannot explain the larger observed impurity capture event yield.

b) Impurity capture events observed in the TPC.

c) Hydrogen flow in the system. Zero describes the situation where the TPC hydrogen vessel is disconnected from CHUPS.

The lowest humidity levels thus far were achieved during the spring 2006 experimental run. For two months before the run, the TPC was continuously vacuum pumped and simultaneously baked at ~120 °C. This procedure led to an initial humidity level of only 60 ppb (Fig. 11) when CHUPS was connected. During 400 hours of continuous cleaning with a mean hydrogen flow rate of 3 slpm, the humidity exponentially decreased to ~18 ppb and remained at this level until the end of the main $\mu^-$ production data taking, providing stable operation over more than 1000 hours. The minor fluctuations of the humidity are explained by temperature fluctuations in the experimental hall. The change of ambient temperature affects the adsorption-desorption equilibrium in the chamber and, consequently, its outgassing rate.

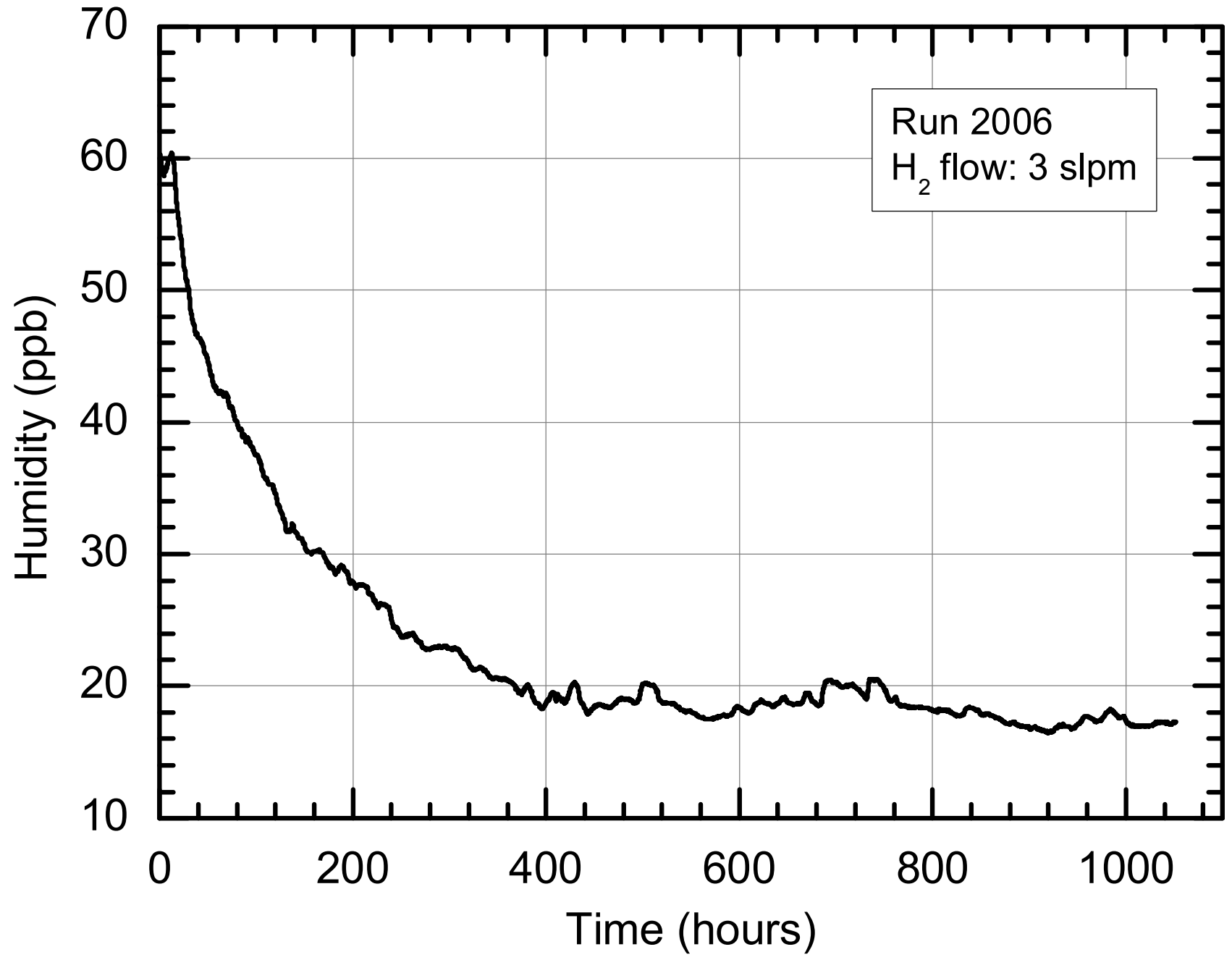

Fig. 11. Humidity decrease in the TPC after connection of the CHUPS system at t=0 to the hydrogen vessel of the TPC as measured with the dew-point transmitter.

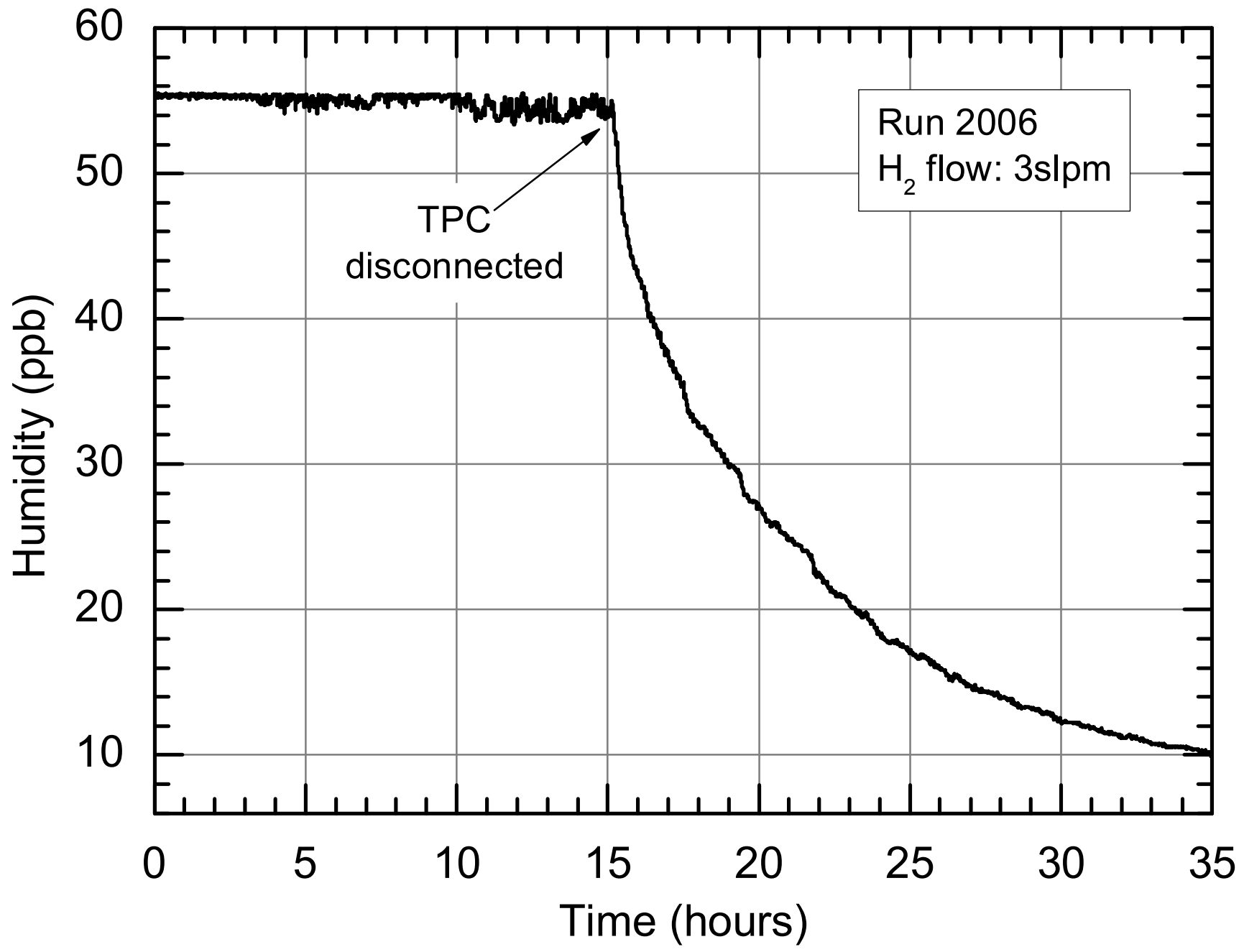

Fig. 12. Humidity change after bypassing the hydrogen vessel as measured with the dew-point transmitter.

Fig. 12 shows the humidity behavior in the CHUPS system after disconnecting the TPC hydrogen vessel, which was determined in a separate test. CHUPS was disconnected from the TPC vessel and run through a bypass line. The humidity decreased exponentially and reached the 10 ppb level within 20 hours. The finally reached CHUPS stand-alone humidity level was 5 ppb as measured in 2005 with the dew-point transmitter.

### 5.2. Impurity doped fills for calibration of the detection efficiency of impurity capture events in the TPC

The cleaning power of CHUPS was confirmed during systematic calibration studies of the TPC detector. Several times, small known amounts of typical impurities were admixed to our hydrogen target. The high cleaning power of the CHUPS system made it possible to return the detector to the normal clean conditions in a short time (1-2 days).

The cleaning efficiency of CHUPS for a large contamination of nitrogen in the detector hydrogen was tested by adding a 22±1 ppm nitrogen admixture produced via dilution in a vessel with high purity hydrogen. This nitrogen-doped condition served as a detection efficiency calibration of nitrogen impurity capture events in the TPC. The nitrogen concentration behavior observed online via capture events in the TPC is shown in Fig. 13. After the measurement with the large nitrogen contamination (flat region) the CHUPS circulation through the TPC vessel was re-established and the nitrogen was removed. The cleaning started with an impurity concentration of more than 1000 ppb and proceeded until leveling off around ~30 ppb, a value established before the doping. The purification process took only 20 hours.

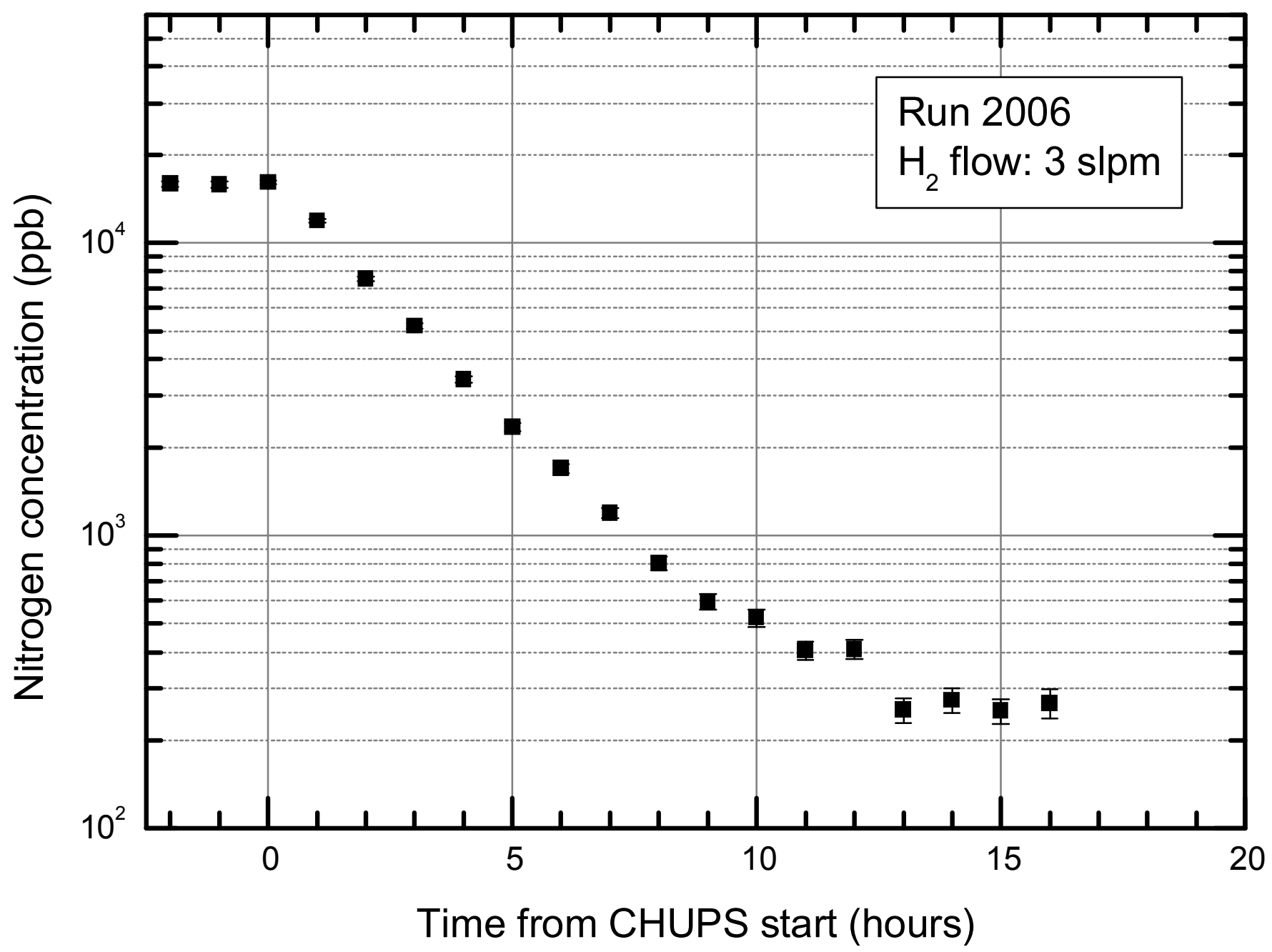

Fig. 13. Decrease of nitrogen concentration as monitored by the TPC impurity capture events in the nitrogen doped run. At t = 0 CHUPS cleaning was started and the impurity capture events, which proportionally track the nitrogen concentration, start to decrease. They reach the previous clean fill level after 13 hours of CHUPS operation.

In another test, a water permeation tube[12] was used to generate high humidity concentrations in the hydrogen flow to calibrate the effect of $H_2O$ impurities. The tube was placed in the temperature stabilized box. Stable hydrogen flow was passed through the volume with the tube. During this calibration experiment the humidity inside the TPC was increased to approximately 2000 ppb. Then CHUPS was connected and the gas was cleaned to 400 ppb in one day (Fig. 14). The slower cleaning speed can be explained by wetting of the inner surfaces of the hydrogen vessel and tubes.

[12] Permeation tube providing 500 ppb ± 10% at 0.5 slpm flow. Valco Instruments Company Inc. P.O. Box 55603, Houston, TX 77255, USA.

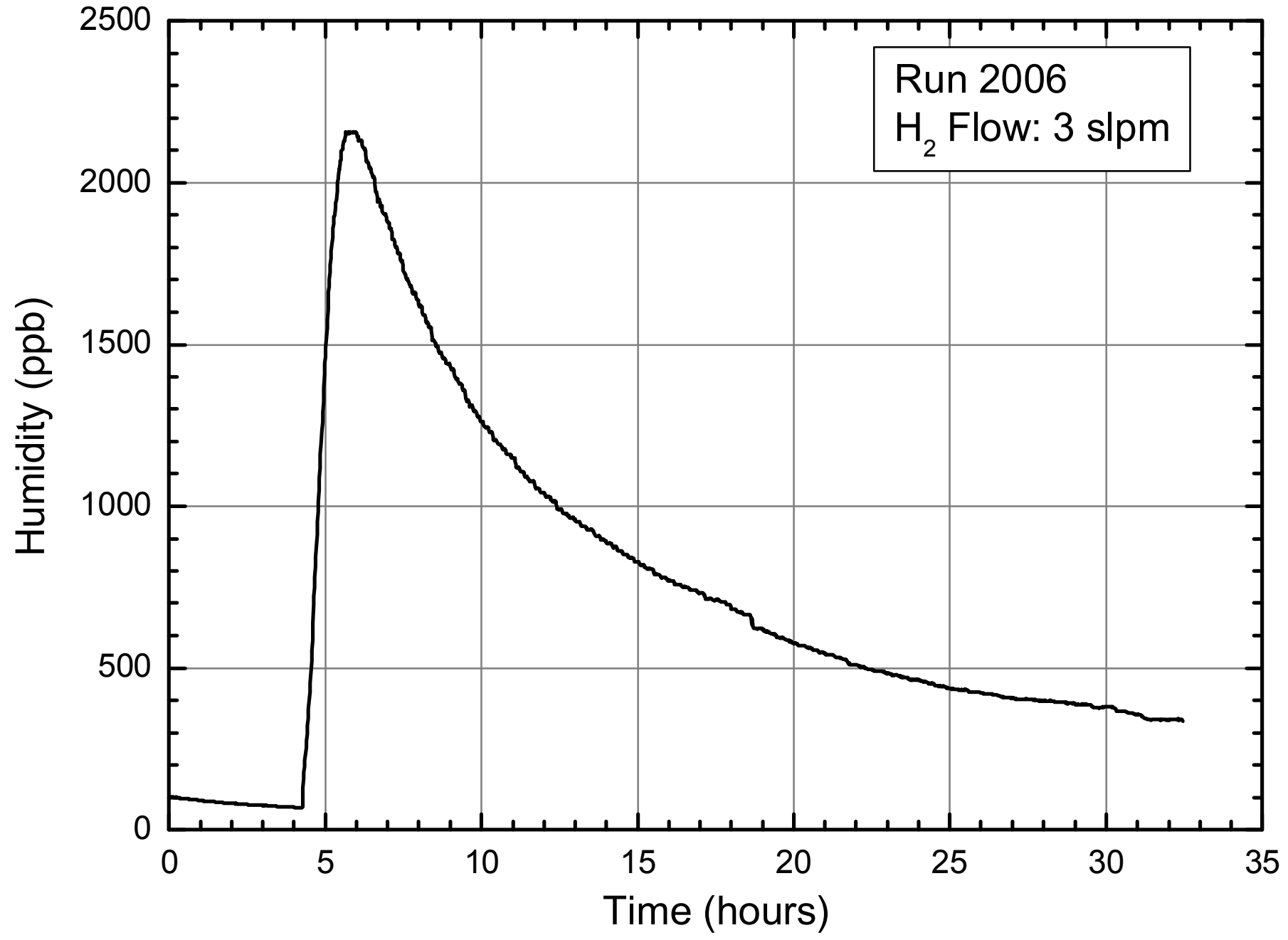

Fig. 14. Drying of TPC during the water doped run as observed with the dew-point transmitter.

In a second test the humidity was kept at the level of about 2 ppm with simultaneous measurements of the dew-point transmitter and the capture yield (Fig. 15). The impurity capture event yields were recalculated to a corresponding humidity level using the factor $k_{H2O}$=205. The gray area in Fig. 15 indicates the uncertainty range of the dew-point transmitter. The impurity capture event yield consistently tracks the dew-point transmitter readings. A fast CHUPS recovery (during hours 38-40) is shown after bypassing the hydrogen vessel. The humidity in the closed detector also changed due to slow moisture adsorption on the walls.

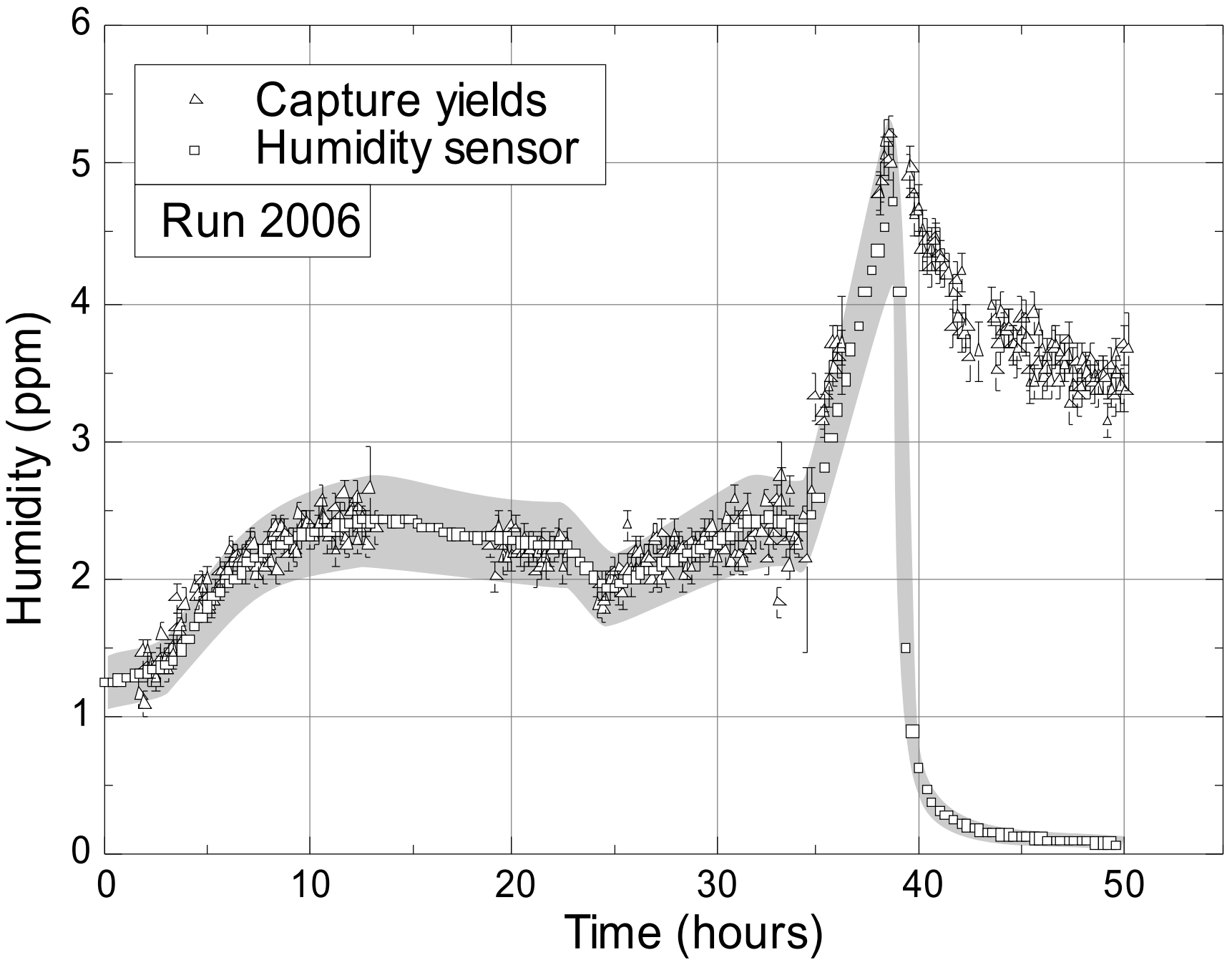

Fig. 15. Simultaneous measurements of dew-point transmitter and of efficiency corrected impurity capture event yield observed in the TPC.

## 6. Results

The CHUPS system was installed in the MuCap experiment at PSI and first connected to the TPC detector located inside a hydrogen-filled pressure vessel in 2004. The system proved to be very reliable and versatile during three experimental runs in 2004-2006.

The performance of the CHUPS system during development and use fully justified the initial design. The main result was a stable clean working gas in the TPC. A smooth hydrogen flow of up to 3 slpm was maintained during the entire TPC operation time (more than 1000 hours in every run). The best purity reached was 18 ppb for moisture, <7 ppb and <5 ppb for nitrogen and oxygen, respectively.

This result is sufficient for the planned precision measurement of the rate $\Lambda_S$ for muon capture on the proton as the remaining impurity level is very low and the small corrections to the negative muon lifetime caused by impurities can be calculated based on calibration measurements.

In addition to regular cleaning, CHUPS provided fast cleaning of the working gas after calibrations with large amounts of contaminants. This feature was essential for systematic studies in specific runs used for calibrating the TPC detection efficiency for nitrogen and water. Variations of the CHUPS flow and correspondingly the equilibrium humidity in the hydrogen vessel gave important additional calibration points for the systematic investigations.

With regard to the stable pressure requirement CHUPS surpassed its design specifications. The pressure inside the TPC was kept at the appropriate level of 10 bar with a stability of 0.024% during all the operation modes over more than 2 months. Pulsations of the hydrogen flow through the TPC vessel were also minimized to the level of 2.5% at a mean flow rate of 3 slpm. The CHUPS system has proven to be of key importance to the MuCap experiment [2] and might find application in other experiments with strict requirements on the gas purity.

## Acknowledgments

The authors wish to thank Prof. A.Vorobyov for important design remarks and fruitful discussions. They gratefully acknowledge the help of the Cryogenic and Superconducting Technique Laboratory and G.A. Ganzha of the Petersburg Nuclear Physics Institute during the development of the system; and the help of Dr. F. Mulhauser during installation and experiments at PSI. We are grateful to the whole MuCap collaboration for their support during the runs and the permission to present MuCap data. This work has been supported by the CRDF under Award No. RP2-2414-GA-02, the US National Science Foundation, PSI and a grant of the President of the Russian Federation (NSH-3057.2006.2).